\documentclass[aps,prl,twocolumn,groupedaddress,showpacs]{revtex4}

\usepackage{graphicx}
\usepackage{amsmath}
\usepackage{amssymb}
\usepackage{bm}

\begin{document}
\title{Resonant control of spin dynamics in ultracold quantum gases by microwave dressing}
\author{Fabrice Gerbier}\email[email: ]{fabrice.gerbier@lkb.ens.fr}
\author{Artur Widera} \author{Simon F{\"o}lling}  \author{Olaf Mandel} \author{Immanuel
Bloch} \affiliation{Institut f{\"u}r Physik, Johannes
Gutenberg-Universit{\"a}t, 55099 Mainz, Germany.}
\date{\today}
\begin{abstract}
We study experimentally interaction-driven spin oscillations in
optical lattices in the presence of an off-resonant microwave
field. We show that the energy shift induced by this microwave
field can be used to control the spin oscillations by tuning the
system  either into resonance to achieve near-unity contrast or
far away from resonance to suppress the oscillations. Finally, we
propose a scheme based on this technique to create a flat sample
with either singly- or doubly-occupied sites, starting from an
inhomogeneous Mott insulator, where singly- and doubly-occupied
sites coexist.

\end{abstract}
\maketitle
%
The fact that coupling an atomic system to a strong radiation
field can modify its level structure is well-known, leading to
such effects as the Autler-Townes doublet or the Bloch-Siegert
shift (see, e.g., \cite{cct_photons_atoms}). Today, this
phenomenon is often explained in a dressed-atom picture, where the
bare atomic states are replaced by  eigenstates of the total
atom-plus-field system. For cold atoms, this picture helps to
understand how magnetic trapping potentials can be engineered with
off-resonant radio-frequency or microwave radiation
\cite{spreeuw1994}, allowing for example to create nearly
two-dimensional \cite{colombe2004a} or double-well trapping
potentials \cite{treutlein2004a,schumm2005a}.

Here, we show how an off-resonant microwave field (hereafter
called {\it dressing field}) can be used for coherent control of
the dynamics of a spinor quantum gas. Ultracold gases with
internal (spin) degrees of freedom \cite{ho1998a} are novel
quantum systems where atom-atom interactions play a key role. In
an optically trapped gas of spin-$f=1$ atoms, spin-dependent
interactions can promote an atom pair from an initial two-atom
state ,where both atoms are initially in the $m=0$ magnetic
sublevel, to a final state with one atom in $m=+1$ and the other
in $m=-1$
\cite{stamperkurn1999b,law1998a,schmaljohann2004a,chang2004a,kuwamoto2004a}.
The coherence of this process was demonstrated in
\cite{widera2005a,widera2005b} for atom pairs in optical lattices
and in \cite{kronjaeger2005a,chang2005a} for a seeded
Bose-Einstein condensate. In the former case, the dynamics is
controlled by the energy mismatch, or detuning, between the
initial and final spin states of the colliding atom pair. This
detuning originates not only from different quadratic Zeeman
shifts in the initial or final two-atom states, but also from
different interaction energies. If positive, as in $^{87}$Rb
\cite{chang2004a,chang2005a,widera2005b}, this ``residual''
interaction detuning prevents to reach the resonance for spin
oscillations by magnetic means.

\begin{figure}
\includegraphics[width=8cm]{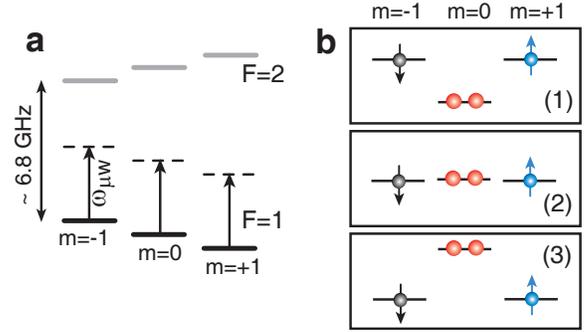}
\caption{ An off-resonant microwave field with approximately
$\pi$-polarization and frequency $\omega_{\mu w}$ ({\bf (a)}) is
used to control the detuning for coherent spin oscillations in
optical lattices. Case {\bf (b1)} corresponds to zero or weak
microwave intensity. With properly chosen microwave parameters,
one can compensate this detuning and achieve full resonance ({\bf
b2}) or overcompensate to suppress the spin-changing collisions
({\bf b3}).)} \label{Fig0}
\end{figure}

In this paper, we show how to tune the spin oscillations in and
out of resonance using the energy shift induced by an off-resonant
dressing field (see Fig.\ref{Fig0}). We apply this technique to an
ensemble of atom pairs in a deep optical lattice and initially
polarized in the $m=0$ sublevel. We show that the conversion to an
ensemble of $m=\pm 1$ pairs can be achieved with near unit
efficiency, even in a finite magnetic field. The significance of
this work goes beyond the topic of spin oscillations in optical
lattices. Controlling the spin oscillations enables one to use
them as a tool for quantum state detection, preparation or
manipulation. In a separate paper, we have described how this
technique can be used to probe atom number statistics across the
superfluid to Mott insulator (MI) transition \cite{gerbier2005c}.
We propose here to use it in order to create an homogeneous MI
with singly-occupied sites out of a inhomogeneous system where
singly- and doubly-occupied sites coexist.

In our case, the microwave is relatively weak, so that the dressed
states almost coincide with the bare atomic states and the level
shifts can be calculated perturbatively \cite{cct_photons_atoms}.
We consider a $^{87}$Rb atom in a constant and homogeneous
magnetic field $B$, plus an arbitrary polarized microwave field
with amplitudes $\{ B^{(\rm \mu w)}_q \}_q$, with $q=0,\pm 1$ for
$\pi$ and $\sigma^\pm$ polarizations, respectively. Far from any
hyperfine resonance, we find the level shift for the state
$|F=1,m\rangle$ as
\begin{eqnarray}\label{levelshift}
\Delta E_{m} \approx \frac{\hbar\Omega_\pi^2}{4\Delta}
f_m\left(x=\frac{\mu_{\rm B} B}{\hbar \Delta} \right).
\end{eqnarray}
In Eq.~(\ref{levelshift}), $\Delta=\omega_{\rm \mu w}-\omega_{\rm
hf}$ denotes the microwave detuning, where $\omega_{\rm \mu w}$ is
the microwave frequency and where $\omega_{\rm hf}$ is the
hyperfine frequency splitting in the $5$S$_{1/2}$ manifold of
Rubidium $87$. We have introduced the Rabi frequency
$\Omega_\pi=\mu_{\rm B} B^{(\rm \mu w)}_\pi/\hbar$ for the
$\pi$-polarized transition, with $\mu_{\rm B}$ the Bohr magneton,
and assumed that $x=\mu_{\rm B} B/\hbar \Delta \ll 1$. The
function
\begin{eqnarray}\label{Eq:f}
f_m(x)=\sum_{q} \frac{3
C_{m,q}}{4}\frac{I_q}{I_\pi}\frac{1}{1-(m+\frac{q}{2}) x},
\end{eqnarray}
accounts for different microwave polarizations and the multi-level
structure. Here $C_{m,q}=| \langle F=1,m;1,q| F=2,m+q \rangle |^2$
are square moduli of Clebsch-Gordan coefficients, and $I_{q}$ is
the intensity of the $q-$polarized microwave component. We
calibrate the microwave intensities for each polarization $q$ by
driving fast, single-particle Rabi flopping between the hyperfine
states $|F=1,m=-1\rangle$ and $|F=2,m=-1+q\rangle$. From this, we
obtain $I_{\sigma^-}/I_\pi\approx 0.33$ and
$I_{\sigma^+}/I_\pi\approx 0.02$ for our experimental parameters.
The peak Rabi frequency observed in our experiment is
$\Omega_\pi=2\pi\times82(1)~$kHz.

\begin{figure}
\includegraphics[width=7.8cm]{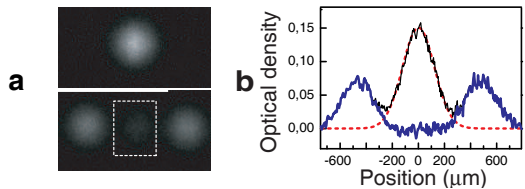}
\caption{{\bf (a)} Absorption image of an expanded cloud after
$t_{\rm osc}=0~$ms (upper image) and $t_{\rm osc}=15~$ms (lower
image). The dashed line indicates the box used to fit the $m=0$
component. {\bf (b)} Horizontal cut through the lower image in
{\bf (a)} (black line). Subtracting a Gaussian fit (red dashed
line) to the $m=0$ component yields a new image (blue line) used
to count separately the population in $m=\pm 1$ by integration. }
\label{Fig1}
\end{figure}
To observe the microwave-induced shift, a MI of around
$2\times10^5$ Rb  atoms is first created in the $|F=1,m=-1\rangle$
Zeeman sublevel. This MI is held in a purely optical lattice
potential at a magnetic field of $1.2~$G. In this work, we use a
typical lattice depth of $40\,E_r$, with $E_r = h^2/2 M \lambda^2$
and $\lambda = 840$\,nm the lattice laser wavelength. After an
experimental phase described below and a 12-ms-long free expansion
including a magnetic gradient pulse, we detect the population in
each Zeeman sublevel as shown in Fig.~\ref{Fig1}
\cite{widera2005a,widera2005b}. We find the population in the
$m=0$ sublevel by fitting a Gaussian function to a restricted
area, by substracting it from the data, and by integrating over
the cloud areas to find the populations in $m=\pm 1$ (see
Fig.~\ref{Fig1}).

The main topic of this paper is the study of spin dynamics in a
dressing field. In our experiments, the spin dynamics is
initialized as described in \cite{widera2005a,widera2005b} by
first transferring the sample to $|F=1,m=0\rangle$ by applying two
microwave $\pi-$pulses starting from $|F=1,m=-1\rangle$. After
these pulses, the magnetic field is eventually ramped down to a
final value (typically $B=0.4\,$G for the experiments described
here). Subsequently, collisional spin oscillations take place at
each individual lattice site where an atom pair is present. As
discussed in \cite{widera2005a,widera2005b}, the atom pair
prepared with both atoms initially in $m=0$ behaves as a two-level
system, and undergoes Rabi-like oscillations at an effective
oscillation frequency
\begin{eqnarray}\label{Eq:Rabi3}
\hbar\Omega_{\rm eff} & = &
\sqrt{(\hbar\Omega_0)^2+(\Delta\epsilon+U_s)^2}.
\end{eqnarray}
Here the coupling strength $\hbar\Omega_{0}=2\sqrt{2}U_s$ and the
interaction part of the detuning are both fixed by a
spin-dependent interaction energy $U_s=\tilde{U}(a_2-a_0)/3$, with
$\tilde{U}$ an on-site interaction energy per Bohr radius $a_B$
\cite{widera2005b} and $a_F$ the scattering length characterizing
scattering in the total spin-$F$ channel
\cite{ho1998a,widera2005a}. The quantity
$\Delta\epsilon=\epsilon_{\rm +1}+\epsilon_{\rm
-1}-2\epsilon_0\propto B^2$ corresponds to the difference in
Zeeman energies $\epsilon_{m}$ between the initial and final
states. Without dressing field, $\Delta\epsilon=(\mu_B B)^2/2\hbar
\omega_{\rm hf}$ is determined by the magnetic field. In the
presence of the dressing field, the Zeeman sublevels shift
according to Eq.~(\ref{levelshift}), so that one obtains
\begin{equation}\label{deltae}
\Delta\epsilon+U_s=\hbar\times\alpha(\Omega_\pi^2-\Omega_{\rm
res}^2),
\end{equation}
where $\alpha=(f_{\rm+1}(x)+f_{\rm -1}(x)-2f_{\rm 0}(x))/4\Delta$
and where the resonance Rabi frequency $\Omega_{\rm res}$ solves
\begin{equation}\label{rescond}
\hbar\alpha\Omega_{\rm res}^2+\frac{(\mu_B B)^2}{2\hbar\omega_{\rm
hf}}+U_s=0.
\end{equation}
When normalized to the total atom number, the oscillation
amplitude for the lattice as a whole is reduced by a factor $n_2$,
corresponding to the fraction of atom pairs in the lattice
\cite{widera2005a,widera2005b}. For our trapping parameters and
our atom number, approximately half of the atoms are in a central
region with two atoms per site ($n_2\approx0.5$)
\cite{widera2005a}.

\begin{figure}
\includegraphics[width=7.8cm]{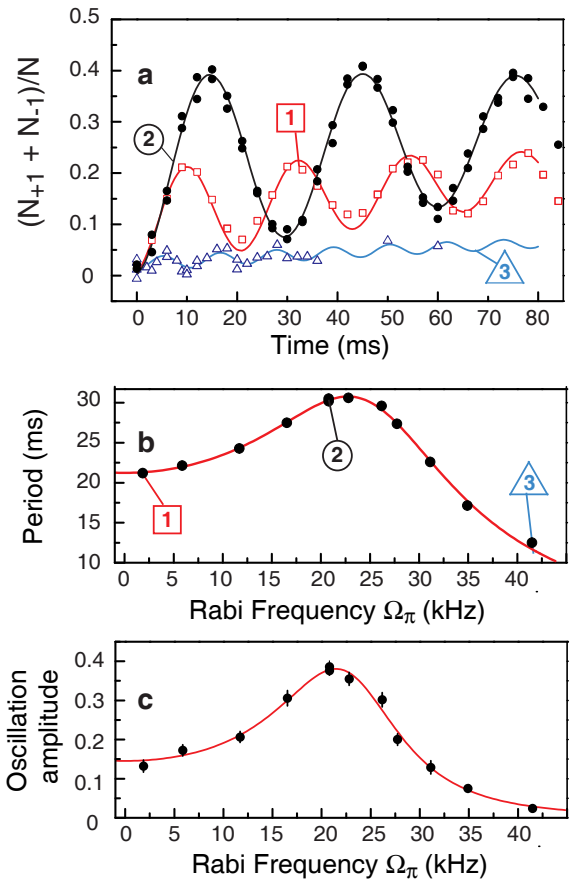}
\caption{Spin oscillations at a magnetic field $B=0.4~$G. {\bf
(a)} Spin oscillations without microwave dressing (squares), for a
resonant dressing (circles) and for an overcompensated dressing
(triangles). {\bf (b)} and {\bf (c)}, period and amplitude of the
oscillation
 (circles), together with a fit to the expected Rabi dependance.} \label{fig2}
\end{figure}

We have investigated the resonantly dressed spin dynamics at a
magnetic field of $0.4~$G, by varying the microwave power (or
equivalently the dressing frequency), and recording several
periods of spin oscillations for each power. The microwave field
was detuned by several hundred MHz to the red of the hyperfine
resonances to suppress population transfer to $F=2$. Three
representative examples are shown in Fig.~\ref{fig2}a,
corresponding to the cases illustrated in Fig.~\ref{Fig0}: the
unperturbed case ($\Omega_\pi\ll\Omega_{\rm res}$, squares), the
near-resonant case ($\Omega_\pi\approx\Omega_{\rm res}$, circles)
where the microwave dressing compensates exactly the ``bare''
detuning, and the overcompensated case ($\Omega_\pi\gg\Omega_{\rm
res}$, triangles), where the dynamics is blocked due to the large
detuning induced by the microwave dressing. We have found
empirically that the data were well described by a fitting
function of the form
\begin{equation} \label{Eq:Rabi2}
    \frac{N_{0}}{N} =
    A \left( 1- e^{-\gamma_{\rm off}t} \right)+
    B \cos \left( 2\pi \frac{t}{T_{\rm osc}} \right) e^{-\gamma_{\rm
    osc}t}.
\end{equation}
The second term describes the spin oscillations with period
$T_{\rm osc}$, including a damping term with rate $\gamma_{\rm
osc}$. The first term describes a slowly varying offset
approaching the value $A$ with a time constant $\gamma_{\rm
off}^{-1}$.
\begin{figure}
\includegraphics[width=7.8cm]{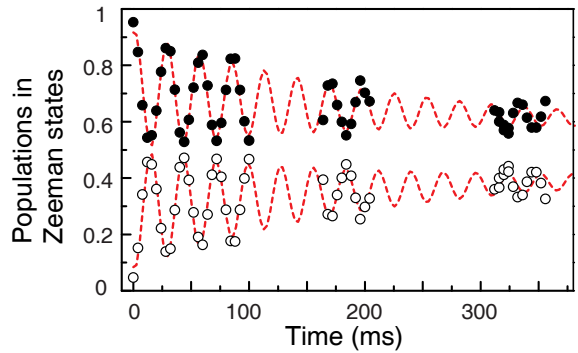}
\caption{Spin oscillations recorded for a magnetic field
$B=1.2$~G. Filled circles denote the population in the Zeeman
substate $m=0$, and hollow circles the sum of the populations in
substates $m=\pm 1$.} \label{fig3}
\end{figure}

The measured period of the oscillation is plotted in
Fig.~\ref{fig2}b as a function of the Rabi frequency $\Omega_\pi$
of the dressing field. The period is fitted to the expected form
of the Rabi frequency given in Eq.~(\ref{Eq:Rabi3},\ref{deltae})
treating $U_s$, $\Omega_{\rm res}$ and $\alpha$ as free
parameters. We find good agreement between the expected
$2\pi\alpha=0.069$\,s, $\Omega_{\rm res}=2\pi\times23.4$\,kHz and
the measured values $2\pi\alpha=0.066(1)$\,s, $\Omega_{\rm
res}=2\pi\times22.8(1)$\,kHz. The ``bare'' Rabi frequency
$U_s/h=10.5(1)$\,Hz is in excellent agreement with an independent
measurement of the spin-dependent collisional coupling, which were
done without a dressing field \cite{widera2005a,widera2005b}.

In Fig.~\ref{fig2}c, we show the oscillation amplitude versus
microwave Rabi frequency $\Omega_\pi$, together with a fit to the
expected behavior. The fit parameters include the same parameters
$\alpha,\Omega_{\rm res}$ as in the previous case, plus an
additional parameter $n_2$ corresponding to the fraction of atoms
participating in the oscillation. The values for $\alpha$ and
$\Omega_{\rm res}$ are consistent with those deduced from the fit
to the measured period. The overall amplitude $n_2\approx0.40(3)$
is slightly smaller than the fraction $n_2=0.5$ of atom pairs
expected in the Mott phase. We have performed another measurement
where the atomic sample is held at a magnetic field of $1.2\,$G
using the dressing field on resonance according to
Eq.~(\ref{rescond}). The spin oscillation curve shown in
Fig.~\ref{fig3} yields an amplitude of $n_2\approx0.47(2)$, much
closer to the expectation than for the $B=0.4\,$G case. We
attribute part of this $\sim15~$\% reduction in contrast to a
finite ramp time of the magnetic field in the latter case,
inducing a small amount of adiabatic population transfer at the
beginning of the oscillation \cite{widera2005b}.
\begin{figure}
\includegraphics[width=7.8cm]{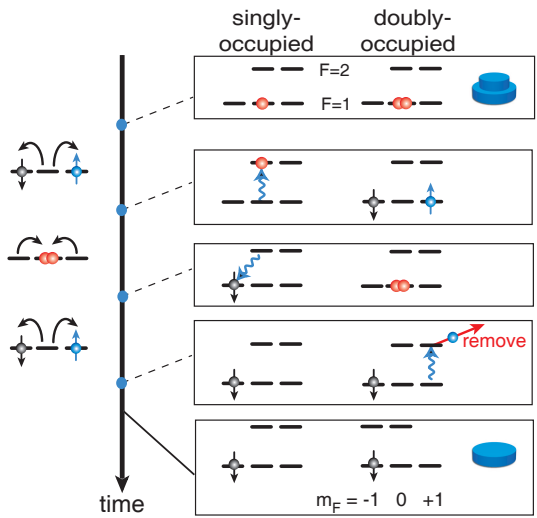}
\caption{Illustration of the proposed scheme to create a
homogeneous MI with only singly-occupied sites starting from a
sample with singly- and doubly-occupied Mott plateaus. After half
a spin oscillation period, atoms in singly-occupied sites are
transferred into the upper hyperfine level. Another half-spin
oscillation brings back atoms in doubly occupied sites to
$|F=1,m=0\rangle$, while a second microwave pulse is used to
transfer atoms in singly-occupied sites to $|F=1,m=-1\rangle$.
Another spin oscillation nearly creates the homogeneous $n=1$
system, the only difference being a supplementary atom in
$|F=1,m=+1\rangle$ in doubly occupied sites. This can be corrected
by transferring these atoms to $F=2$, where a resonant laser beam
can be used to remove them out of the trap.} \label{fig4}
\end{figure}
From the data shown in Fig.~\ref{fig3}, we have also determined
the damping times $\gamma_{\rm osc},\gamma_{\rm off}$ and the
offset value $A$, as defined in Eq.~\ref{Eq:Rabi2}. Within our
experimental accuracy, we find that they are independent of
microwave power and magnetic field. The average offset value is
$A\approx0.6$, and the average damping times $\gamma_{\rm
osc}\approx140$\,s$^{-1}$ and $\gamma_{\rm
off}\approx160$\,s$^{-1}$. Although the origin of the damping is
not entirely clear, the fact that $\gamma_{\rm
osc}\approx\gamma_{\rm off}$ indicates a correlation between the
damping of the oscillations and the rise of the offset. Possible
damping mechanisms can be either local ({\it i.e.} involving only
on-site processes) or non-local (involving tunneling to a
neighboring site, particularly effective if atoms are transferred
to excited Bloch bands by heating processes). On-site damping
mechanisms could possibly be off-resonant Raman transitions
between the $m=\pm 1$ sublevels induced by the lattice beams. We
believe the limitations of the present experiment to be of
technical origin. In principle, even longer coherence times could
be achieved.

Because of these long coherence times, the technique presented
here could find applications beyond the study of spin
oscillations. For example, Refs.
\cite{pu2000a,duan2000a,sorensen2001a} discuss how spin squeezing
experiments could benefit from such a control. In
Fig.\,\ref{fig4}, we show how a homogeneous MI with one atom per
site only can be created, starting from of an inhomogeneous system
where Mott plateaus with $n=1,2$ coexist. This could be important
for the implementation of parallel collisional quantum gates in a
lattice \cite{jaksch1999a,mandel2003b}, where an inhomogeneous
filling degrades the fidelity of the process (see also
\cite{rabl2003a}). Similarly, one can produce an array of
doubly-occupied sites in $m=-1$, as first proposed in
\cite{widera2004a} (see also \cite{widera2005a,thalammer2005a}).

In conclusion, we have demonstrated resonant tuning of spin
oscillations in optical lattices using an off-resonant microwave
field. This provides dynamical control over the spin oscillations,
which together with the long coherence times enables to use them
as a tool for detection, preparation and manipulation of the
quantum state of the many-atom system. As an illustration, we
propose a filtering technique producing a flat MI with a single
atom per site, starting from a system with singly- and
doubly-occupied sites. From a thermodynamical point of view, this
flat MI would be metastable with a lifetime on the order of the
tunneling time in the system. This method could also be applied to
a thermodynamically stable system with only singly-occupied sites
in the ground state. This would help to remove local defects with
two atoms on a site, thus providing a novel cooling technique
unique to the MI state.
\begin{acknowledgments}
We acknowledge support from the DFG, from AFOSR and from the EU
under the OLAQUI program.
\end{acknowledgments}
%

%
\end{document}